\documentclass{biometrika}

\usepackage{amsmath,amsfonts, amssymb} 
\usepackage{times}

\usepackage[pdftex]{graphicx}
\usepackage{natbib}

\begin{document}

\title{Adaptive approximate Bayesian computation}

\author{Mark A. Beaumont}\affil{School of Biological Sciences, University of Reading,
PO Box 68, Whiteknights, Reading  RG6 6BX, U.K.
\email{m.a.beaumont@reading.ac.uk}}
\author{Jean-Marie Cornuet}\affil{Department of Epidemiology and Public Health, Imperial College,
London SW7 2AZ, U.K. \email{jmcornuet@ensam.inra.fr}}
\author{Jean-Michel Marin}\affil{Institut de Math\'ematiques et Mod\'elisation de Montpellier,
Universit\'e Montpellier 2, Case Courrier 51, 34095 Montpellier cedex 5, France \email{Jean-Michel.Marin@univ-montp2.fr}}
\author{Christian P.~Robert}\affil{Centre de Recherche en Math\'ematiques de la D\'ecision, 
Universit\'e Paris Dauphine, 75775 Paris cedex 16, France \email{xian@ceremade.dauphine.fr}}
	
\maketitle
\markboth{Beaumont, Cornuet, Marin, and Robert}{Adaptive approximate Bayesian computation}

\begin{abstract}
Sequential techniques can enhance the efficiency of the approximate Bayesian computation algorithm, as in
Sisson et al.'s (\citeyear{sisson:fan:tanaka:2007}) partial rejection control version. While this 
method is based upon the theoretical works of \cite{delmoral:doucet:jasra:2006}, the application 
to approximate Bayesian computation results in a bias in the approximation to the posterior.
An alternative version based on genuine importance sampling arguments bypasses this difficulty,
in connection with the population Monte Carlo method of \cite{cappe:guillin:marin:robert:2003}, 
and it includes an automatic scaling of the forward kernel. When applied to a population genetics 
example, it compares favourably with two other versions of the approximate algorithm.

\end{abstract}

\begin{keywords}Markov chain Monte Carlo; partial rejection control; importance sampling; sequential Monte Carlo. 
\end{keywords}

\section{Introduction}

When the likelihood function is not available in a closed form, as in population genetics,
\cite{pritchard:seielstad:perez:feldman:1999} introduced approximate Bayesian computational methods 
as a rejection technique bypassing the computation of the likelihood function
via a simulation from the corresponding distribution. Namely, if we observe $y\sim f(y\mid\theta)$
and if $\pi(\theta)$ is the prior distribution on the parameter $\theta$, then the 
original approximate Bayesian computation algorithm jointly simulates
$\theta^\prime\sim \pi(\theta)$ and $x\sim f(x\mid\theta^\prime)$,
and accepts the simulated $\theta^\prime$ if, and only if, the auxiliary variable $x$ is equal to the observed value,
$x=y$. This algorithm is exact in that the accepted $\theta^\prime$'s are distributed from the
posterior. In the more standard occurrence when $y$ is a continuous random variable, the approximate Bayesian computation algorithm 
relies on an
approximation, where the equality $x=y$ is replaced with a tolerance condition, $\varrho(x,y)\le\epsilon$, $\varrho$
being a measure of discrepancy, for instance a distance between summary statistics, and $\epsilon>0$ being the tolerance
bound. The output is then distributed from the distribution with density proportional to 
	$\pi(\theta)\,\text{pr}_\theta\{\varrho(x,y)<\epsilon\}$,
where $\text{pr}_\theta$ represents the probability distribution of $x$ indexed by a value of $\theta$. This density
is denoted by $\pi\{\theta\mid\varrho(x,y)<\epsilon\}$, where the conditioning corresponds to
the marginal distribution of $\varrho(x,y)$ given $y$. Improvements to this general scheme have
currently been achieved either by modifying the proposal distribution of the parameter $\theta$ to increase the density 
of $x$'s within a neighbourhood of $y$ \citep{marjoram:etal:2003,bortot:coles:sisson:2007}
or by viewing the problem as a conditional density estimation and developing techniques to allow for larger $\epsilon$
\citep{beaumont:zhang:balding:2002}.

\cite{marjoram:etal:2003} defined a Markov chain Monte Carlo version of the approximate Bayesian computation algorithm
that enjoys the same validity as the original algorithm, namely that, if a Markov chain $(\theta^{(t)})$ is created via the transition 
that sets $\theta^{(t+1)}$ equal to a new value $\theta^\prime\sim K(\theta^\prime\mid\theta^{(t)})$ only if $x\sim f(x\mid\theta^\prime)$
is such that $x=y$ and if $u\sim\mathcal{U}(0,1) \le {\pi(\theta^\prime)
                K(\theta^{(t)}\mid\theta^\prime)}\big/{\pi(\theta^{(t)}) K(\theta^\prime\mid\theta^{(t)})}$,
then its stationary distribution is the posterior distribution $\pi(\theta\mid y)$. Since, in most settings,
the distribution of $y$ is absolutely continuous, the above constraint $x=y$ is replaced with the approximation
$\varrho(x,y)<\epsilon$.

\begin{example}\label{ex:toy2.0}
For the toy model studied  in  \cite{sisson:fan:tanaka:2007}, where
$\theta\sim\mathcal{U}(-10,10)$ and $x\mid \theta\sim 0\cdot5\,\mathcal{N}(\theta,1) + 0\cdot5\,\mathcal{N}(\theta,1/100)$,
the posterior distribution associated with the observation $y=0$ is the normal mixture
$$
\theta\mid y=0\sim 0\cdot5\,\mathcal{N}(0,1)+ 0\cdot5\,\mathcal{N}(0,1/100)
$$
restricted to the set $[-10,10]$. As in regular Markov chain Monte Carlo settings, the performance of 
Marjoram et al.'s (2003) algorithm depends on the choice of the scale $\tau$ in the random walk proposal, 
$K(\theta^\prime\mid \theta)=\tau^{-1}\varphi\{\tau^{-1} (\theta-\theta^\prime)\}$, where $\varphi$ is most often
a standardised normal or a $t$ density. 
However, even when $\tau=0\cdot15$ as in \cite{sisson:fan:tanaka:2007}, the Markov chain mixes slowly, 
but still produces an acceptable fit over $T=10^6$ iterations.
Furthermore, the true target is available here as
\[
\pi(\theta\vert\, \mid x\mid <\epsilon) \propto \Phi(\epsilon-\theta)-\Phi(-\epsilon-\theta)+
\Phi\{10(\epsilon-\theta)\}-\Phi\{-10(\epsilon+\theta)\}
\]
and, for the value $\epsilon=0\cdot025$, it is indistinguishable
from the exact posterior density $\pi(\theta\mid y=0)$. We can thus clearly separate 
the issue of poor convergence of the algorithm, depending on $\tau$, 
from the issue of approximating the posterior density, depending on $\epsilon$.
\end{example}

\cite{sisson:fan:tanaka:2007} modified the approximate Bayesian computation
algorithm using partial rejection control, introduced in \cite{liu:2001}.  The method 
is sequential in that simulated populations of $N$ points are generated at each iteration of the algorithm and that 
those populations are exploited to produce better proposals for a given target distribution in later iterations. As demonstrated in
\cite{douc:guillin:marin:robert:2005}, the reliance on earlier populations to build proposals preserves
convergence properties provided an importance sampling perspective is adopted. 
We also recall that a progressive improvement in the performance of proposals is the appeal of using a sequence 
of samples, rather than a single one. The sequential feature of the problem is augmented by the fact that
the tolerance $\epsilon$ is decreasing with iterations, being chosen either from a deterministic 
scale or from quantiles of earlier iterations as in \cite{beaumont:zhang:balding:2002}.

Sisson et al.'s (2007) algorithm produces samples $(\theta_1^{(t)}, \cdots,\theta_N^{(t)})$ by using, at iteration $t=1$,
a regular approximate Bayesian computation step and, at each iteration $t=2,\ldots,T$,
Markov transition kernels $K_t$ for the generation of the $\theta_i^{(t)}$'s, namely 
	$\theta_i^{(t)} \sim K_t(\theta\mid \theta^\star)\,,$
until $x\sim f(x\mid \theta_i^{(t)})$ is such that $\varrho(x,y)<\epsilon$,
where $\theta^\star$ is selected at random among the previous $\theta_i^{(t-1)}$'s with probabilities 
$\omega_i^{(t-1)}$ $(i=1,\ldots,N)$. The probability $\omega_i^{(t)}$ is derived by an importance 
sampling argument,
\begin{equation}\label{eq:sissonPRC22}
\omega_i^{(t)} \propto {\pi(\theta_i^{(t)}) L_{t-1}(\theta^\star\mid \theta_i^{(t)})}
\{\pi(\theta^\star)K_t(\theta_i^{(t)}\mid \theta^\star)\}^{-1}\,,\
\end{equation}
where $L_{t-1}$ is an arbitrary transition kernel. In their examples, \cite{sisson:fan:tanaka:2007} use
$L_{t-1}(\theta^\prime\mid \theta)\allowbreak=K_t(\theta\mid \theta^\prime)$, which means
that the weights are all equal under a uniform prior. The ratio \eqref{eq:sissonPRC22} is inspired from 
\cite{delmoral:doucet:jasra:2006} who use a sequence of backward kernels $L_{t-1}$ in a sequential Monte Carlo 
algorithm to achieve unbiasedness, up to the renormalisation effect,
in the marginal distribution of the current value without computing this intractable marginal.

In this paper, we show via both theoretical and experimental arguments that the weight \eqref{eq:sissonPRC22}
is biased.  Moreover, we introduce a population Monte Carlo correction to the algorithm.
It is based on genuine importance sampling 
arguments and we demonstrate its applicability as well as the improvement it brings compared with the partial rejection control
version. We also illustrate its efficiency in a population genetics example, when compared with two standard alternatives.

\section{Bias of the partial rejection control version}\label{sec:bias}

\subsection{Distribution of the partial rejection control sample}
In order to establish the bias in the weight \eqref{eq:sissonPRC22} as clearly as possible,
we consider the limiting case when $\epsilon=0$. In that case, both Pritchard et al.'s (1999) and Marjoram et al.'s (2003)
algorithms are correct samplers from $\pi(\theta\mid y)$. 
The corresponding partial rejection control version selects a $\theta^{(t-1)}$ from the previous sample and then generates 
both $\theta^\prime\sim K_t(\theta\mid \theta^\star)$ and $x\sim f(x\mid \theta^\prime)$ until $x=y$. 
To properly evaluate the bias associated with one step of Sisson et al.'s (2007) algorithm, 
we also assume that the previous sample is correctly generated from the target, i.e.~that $\theta^\star\sim \pi(\theta\mid y)$. 
Then, denoting by $\theta^{(t-1)}$ the selected $\theta^\star$,
the joint density of the accepted pair $(\theta^{(t-1)},\theta^{(t)})$ is proportional to
$
  \pi(\theta^{(t-1)}\mid y) K_t(\theta^{(t)}\mid \theta^{(t-1)}) f(y\mid \theta^{(t)}),
$
where the normalisation constant only depends on $y$. Using \eqref{eq:sissonPRC22},
the weighted distribution of $\theta^{(t)}$ is such that, for an arbitrary integrable function $h(\theta)$,
$E\{\omega_t h(\theta^{(t)})\}$ is proportional to
\begin{align*}
\iint &h(\theta^{(t)})\, \frac{\pi(\theta^{(t)})L_{t-1}(\theta^{(t-1)}\mid \theta^{(t)})}{\pi(\theta^{(t-1)})
K_t(\theta^{(t)}\mid \theta^{(t-1)})}\, \pi(\theta^{(t-1)}\mid y) K_t(\theta^{(t)}\mid \theta^{(t-1)}) f(y\mid \theta^{(t)})
\text{d}\theta^{(t-1)} \text{d}\theta^{(t)} \\
&\propto
\iint h(\theta^{(t)})\, \frac{\pi(\theta^{(t)})L_{t-1}(\theta^{(t-1)}\mid \theta^{(t)})}
{\pi(\theta^{(t-1)})K_t(\theta^{(t)}\mid \theta^{(t-1)})}
\pi(\theta^{(t-1)}) f(y\mid \theta^{(t-1)}) \\
&\qquad\times 
K_t(\theta^{(t)}\mid \theta^{(t-1)}) f(y\mid \theta^{(t)})
\text{d}\theta^{(t-1	)} \text{d}\theta^{(t)} \\
&\propto \int h(\theta^{(t)}) \pi(\theta^{(t)}\mid y)
\left\{ \int L_{t-1}(\theta^{(t-1)}\mid \theta^{(t)}) f(y\mid \theta^{(t-1)})
\text{d}\theta^{(t-1)}\right\}\, \text{d}\theta^{(t)}\,
\end{align*}
with all proportionality terms being functions of $y$ only. Therefore we can conclude that 
there is a bias in the weight \eqref{eq:sissonPRC22}
unless the inner function integrates in $\theta^{(t-1)}$ to the same constant for all values of $\theta^{(t)}$. 
Apart from this special case, which is achievable when $L_{t-1}(\theta^{(t-1)}\mid \theta^{(t)}) =
g(\theta^{(t-1)})$ but not in the random 
walk type proposal when $L_{t-1}(\theta^{(t-1)}\mid \theta^{(t)})= \varphi(\theta^{(t-1)}-\theta^{(t)})$, 
\eqref{eq:sissonPRC22} is incorrect since the weighted output is not distributed from $\pi(\theta\mid y)$.  

Paradoxically, the weight \eqref{eq:sissonPRC22} used in this partial rejection control version 
misses a $f(y\mid \theta^{(t-1)})$ term in its denominator, 
while the method is used when $f(y\mid \theta)$ is not available. This is exactly the difference between 
the weights used in \cite{sisson:fan:tanaka:2007} and those used in \cite{delmoral:doucet:jasra:2006}, namely 
that, in the latter paper, the posterior $\pi(\theta^{(t-1)}\mid y)$ explicitly appears in the denominator 
instead of the prior. The accept-reject principle at the core of approximate Bayesian computation 
allows for the replacement of the posterior by the prior in the numerator of the ratio \eqref{eq:sissonPRC22}, 
but not in the denominator.

\subsection{A mixture illustration}
\begin{figure}
\centerline{\includegraphics[width=.99\textwidth]{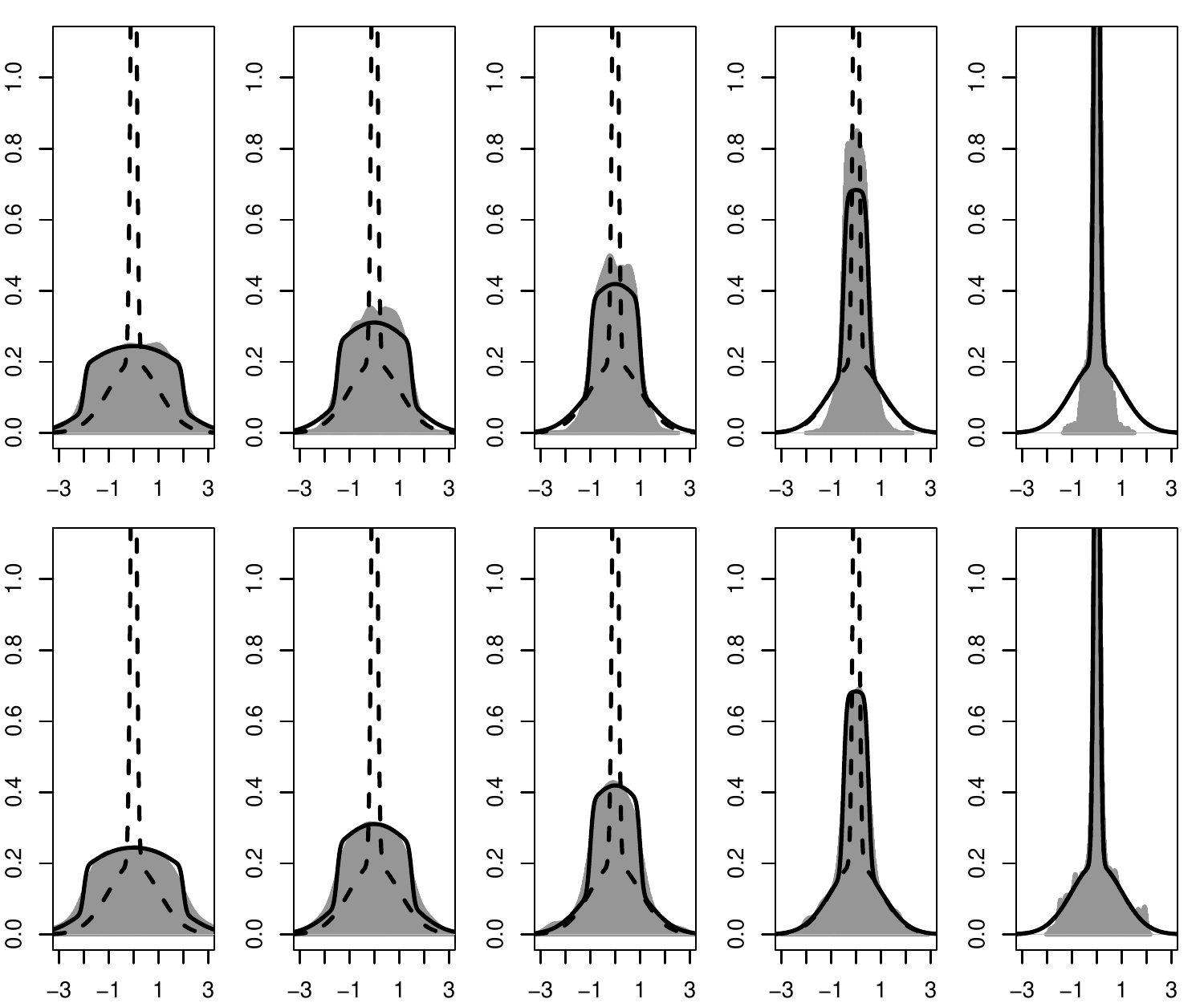}}
\caption{
Histograms of the weighted samples of size $M=5\times10^3$ produced by five consecutive uses of 
\eqref{eq:sissonPRC22} on the first row and the population Monte Carlo version on the second row, 
for the mixture target,
with $\epsilon_t=2,1.5,1,0\cdot5,0\cdot01$, using for both rows the scale $\tau=0\cdot15$.
The exact posterior density is plotted as a dotted curve, while the target of the simulation 
algorithm, $\pi\{\theta\mid \varrho(x,y)<\epsilon\}$, is represented by full lines. Both densities are
identical in the last column.}\label{fig:sismix}
\end{figure}

When using the standard version of the algorithm that includes an
additional approximation due to the tolerance zone $\varrho(x,y)<\epsilon$, there is no reason for the 
bias to vanish, even though experiments show that the bias in the weights \eqref{eq:sissonPRC22} for the tolerance target
$\pi\{\theta\mid \varrho(x,y)<\epsilon\}$ generally decreases as $\epsilon$ increases, which agrees with the
fact that the limiting case is the prior. The following example illustrates this point:

\setcounter{example}{0}
\begin{example}\label{ex:toy2}
For the mixture target 
and a normal random walk kernel $K_t$, the first row of Figure \ref{fig:sismix} 
shows the output of five consecutive iterations of Sisson et al.'s (2007) algorithm, using a decreasing sequence of 
$\epsilon_t$'s, from $\epsilon_1=2$ down to $\epsilon_{5}=0\cdot01$, and a standard deviation in $K_t$ equal to $\tau=0\cdot15$. 
Clearly, using this algorithm leads to a bias in terms of the tolerance target for all values of $\epsilon_t$, 
the tails being poorly covered. In contrast, using a much larger $\tau=1/0\cdot15$ results in
a good fit of the tolerance target, as shown by Figure 2 in \cite{sisson:fan:tanaka:2007}. This fit does
not contradict the bias exhibited above since using a large scale in the proposal 
$K_t$ very closely amounts to using a flat prior distribution. The corresponding normal density is then 
almost constant in the interval $[-3,3]$ that supports the posterior distribution. For $\tau=1/0\cdot15$, using the
weight \eqref{eq:sissonPRC22} is thus equivalent to Pritchard et al.'s (1999) 
original algorithm and therefore as inefficient for this target.
\end{example}

\section{Correction via importance sampling and population Monte Carlo}\label{sec:cor}
\subsection{Population Monte Carlo}
Since the missing factor in \eqref{eq:sissonPRC22}
is the unknown likelihood $f(x\mid \theta^{(t-1)})$, a first resolution of the problem is to resort to an estimation of 
the likelihood based on earlier samples. But a standard importance sampling perspective allows for a 
more direct approach, in a spirit similar to the population Monte Carlo algorithm of \cite{cappe:guillin:marin:robert:2003}.

Since the $t$-th iteration sample 
is produced from the proposal distribution
\[
\hat\pi_t(\theta^{(t)}) \propto \sum_{j=1}^N \omega^{(t-1)}_j K_t(\theta^{(t)}\mid \theta^{(t-1)}_j)\,,
\]
a natural importance weight associated with an accepted simulation $\theta^{(t)}_i$ is
\[
\omega^{(t)}_i \propto {\pi(\theta^{(t)}_i)}\big/{\hat\pi_t(\theta^{(t)}_i)}\,.
\]
The unbiasedness of this correction results from the identity
\[
{E}\{\omega^{(t)} h(\theta^{(t)})\} \propto \int h(\theta^{(t)})\, 
\frac{\pi(\theta^{(t)})}{\hat\pi(\theta^{(t)})}\,
\hat\pi(\theta^{(t)}) f(\theta^{(t)}\mid y) \tilde\pi(\mathbf{\theta}^{(t-1)})\,
\text{d}\theta^{(t)}\,\text{d}\mathbf{\theta}^{(t-1)}\,,
\]
which does not depend on the distribution $\tilde\pi$ of the $\theta^{(t-1)}_j$'s.
As in the original population Monte Carlo method,
the fact that $K_t$ may depend on
simulations from earlier iterations does not jeopardise the validity of the method. In addition, 
\cite{douc:guillin:marin:robert:2005} proved that $K_t$ must be modified at each iteration for the 
iterations to bring an asymptotic improvement on the Kullback--Leibler
divergence between the proposal $\hat\pi_t$ and the target: 
for instance, if the variance of the random walk does not change from one iteration to the 
next, the approximation of the target by $\hat\pi_t$ does not change either 
and it is then more efficient to run a single iteration with twice as many points.
Since $\hat\pi_t$ is an approximation to the distribution of the 
sample simulated at iteration $t$, when marginalised against all previous samples, this version of
the approximate Bayesian computation algorithm can be interpreted as an approximate version of the 
sequential Monte Carlo algorithm of \cite{delmoral:doucet:jasra:2006}, 
when using the optimal backward kernel.

When considering component-wise independent random walk proposals,
\[
K_t(\theta_k^{(t)}\mid \theta_k^{(t-1)})=\tau_k^{-1}\varphi\{\tau_k^{-1}(\theta_k^{(t)}-\theta_k^{(t-1)})\}
\]
for each component $\theta_k^{(t)}$  of the parameter vector $\theta^{(t)}$, 
the asymptotically optimal choice of the scale factor $\tau_k$ is available.
Indeed, when using a Kullback--Leibler measure of divergence between the target and the proposal,
\[
{E}\left( \log \left[ {\pi(\theta^{(t)}\mid y)}\bigg/{\prod_k 
\tau_k^{-1}\varphi\{\tau_k^{-1}(\theta_k^{(t)}-\theta_k^{(t-1)})\}\,f(y\mid \theta^{(t)})} \,,
\right]\right)
\]
where the expectation ${E}$ is taken under the product distribution 
$
(\theta^{(t)},\theta^{(t-1)}) \sim \pi(\theta^{(t)}\mid y)\times\pi(\theta^{(t-1)}\mid y)\,,
$
the minimisation of the Kullback divergence leads to the  component-wise maximisation of
$
{E}[ \log \tau_k^{-1}\varphi\{\tau_k^{-1}(\theta_k^{(t)}-\theta_k^{(t-1)})\}]
$
under the product distribution $\pi(\theta^{(t)}\mid y)\times\pi(\theta^{(t-1)}\mid y)$.  The optimal scale is
then equal to ${E}\{(\theta_k^{(t)}-\theta_k^{(t-1)})^2\}$, that is,
$
\tau_k^2 = 2 \text{var}(\theta_k\mid y)\,,
$
under the posterior distribution. The implementation of this updating scheme on the scale is straightforward. 

The algorithmic rendering of the corresponding optimised population Monte Carlo scheme is thus as follows:\\
\noindent{\em Population Monte Carlo approximate Bayesian computation:}
\begin{quote}
{\small
\noindent Given a decreasing sequence of tolerance thresholds $\epsilon_1\ge\cdots\ge\epsilon_T$,
\begin{itemize}
\item[1.] At iteration $t=1$,
\begin{itemize}
  \item[] For $i=1,...,N$
    \begin{itemize}
    \item[ ]Simulate $\theta_i^{(1)}\sim \pi(\theta)$ and $x\sim f(x\mid \theta_i^{(1)})$ until $\varrho(x,y)<\epsilon_1$
    \item[ ]Set $\omega^{(1)}_i=1/N$
    \end{itemize}
\end{itemize}
Take $\tau^2_1$ as twice the empirical variance of the $\theta_i^{(1)}$'s
\item[2.] At iteration $2\le t\le T$,
\begin{itemize}
  \item[] For $i=1,...,N$, repeat
    \begin{itemize}
    \item[] Pick $\theta_i^\star$ from the $\theta_j^{(t-1)}$'s with probabilities $\omega_j^{(t-1)}$\\
      generate $\theta_i^{(t)}\mid \theta_i^\star \sim \mathcal{N}(\theta_i^\star,\tau_{t}^2)$
      and $x\sim f(x\mid \theta_i^{(t)})$
    \end{itemize}
    until $\varrho(x,y)<\epsilon_t$\\
    Set $\omega^{(t)}_i\propto \pi(\theta^{(t)}_i)/\sum_{j=1}^N \omega^{(t-1)}_j
\varphi\left\{ \tau_t^{-1}\left(\theta^{(t)}_i-\theta^{(t-1)}_j\right)\right\}$
\end{itemize}
Take $\tau_{t+1}^2$ as twice the weighted empirical variance of the $\theta_i^{(t)}$'s
\end{itemize}
}\end{quote}

The expression of the importance weight $\omega^{(t)}_i$ involves a sum of $N$ terms and, therefore, the computational
cost of this step of the population Monte Carlo approximate Bayesian computation algorithm is in $\text{O}(T\,N^2)$. However,
in realistic applications of approximate Bayesian computations, the main cost is associated with the previous step, namely
the repeated simulations from the sampling density. For instance, in the population genetics example, the algorithm spends
at least $95\%$ of the computing time in the repeat loop and less than $5\%$ of the time on the remaining computations.

\subsection{A mixture illustration}

\setcounter{example}{0}
\begin{example}
For the mixure target,
using the population Monte Carlo version leads to a recovery of the target, whether using a fixed standard deviation 
$\tau=0\cdot15$ as shown on the second row of Figure \ref{fig:sismix}, or $\tau=1/0\cdot15$, or a sequence of adaptive 
$\tau_t$'s as in the above algorithm.  The graphical difference between these implementations is genuinely 
difficult to spot; the estimated variance stabilises very quickly in this toy example. 
\end{example}

\subsection{A population genetics example}

This example considers a simple evolutionary scenario of two populations having diverged from a 
common ancestral population. Data consists of the genotypes at five microsatellite loci of 50 diploid
individuals from each of the populations. Loci are assumed to evolve
according to the strict stepwise mutation model: when a
mutation occurs, the number of repeats of the mutated gene increases or
decreases by one unit with equal probability. Once diverged, populations
do not exchange gene and there is no migration. The natural parameters of this
model are the three effective population sizes, $N_1$, $N_2$
and $N_{anc}$, the time of divergence ($t_{div}$) and the mutation rate
($\mu$) assumed here to be common to all loci. While the analyses will
be performed with these natural parameters, only identifiable
combinations of those, such as the three parameters 
$\theta_1=4N_1\mu$, $\theta_2=4N_2\mu$ and $\theta_{a}=4N_{anc}\mu$, and
the parameter $\tau_{div}=t_{div}\mu$, will be considered in the output.

In this experiment, several simulated datasets have been produced using the 
software developed by \cite{cornuet:santos:beaumont:etal:2008},
with the following parameter values: $N_1=N_{anc}=10,000$, $N_2=2,000$, $t_{div}=1,000$ 
and $\mu=0\cdot0005$, out of which one is presented in this paper, but
identical conclusions were drawn from the other datasets.

The tolerance region $\{\varrho(x,y)<\epsilon\}$ used
in the approximate Bayesian computation schemes is based on twelve summary statistics
as in Cornuet et al. (2008), namely
mean number of alleles, mean genic diversity, mean size variance and mean
Garza-Williamson $M$ index for each population sample, $F_{ST}$ and
$(\delta\mu)^2$ distances between population samples, and mean probability
of assignment of each sample to the other population. The distance $\varrho(x,y)$
is then chosen as the Euclidean distance between the observed and the simulated
summary statistics,  normalised by their standard deviation under the predictive model.
The derivation of the distance thus requires a preliminary evaluation of those standard deviations 
for each summary statistic by simulation. 

Each dataset has been submitted to three parallel analyses: a standard approximate Bayesian computation
analysis following \cite{beaumont:zhang:balding:2002},
performed via Cornuet et al.'s (2008) software, a tempered Markov chain Monte Carlo analysis as described in 
\cite{bortot:coles:sisson:2007}, and a population Monte Carlo analysis. In each case, the same prior distributions 
have been used, consisting in a $U[10^2, 10^5]$ prior on the three effective
sizes, a $U[10,10^4]$ prior on the time of divergence, and a $U[10^{-4},10^{-3}]$
prior on the mutation rate. In addition, since Beaumont et al.'s (2002) algorithm 
includes a final local regression adjustment, this
adjustment has been performed on the output of both alternative algorithms. 

\begin{figure}
\centerline{\includegraphics[scale=.7,width=.9\textwidth]{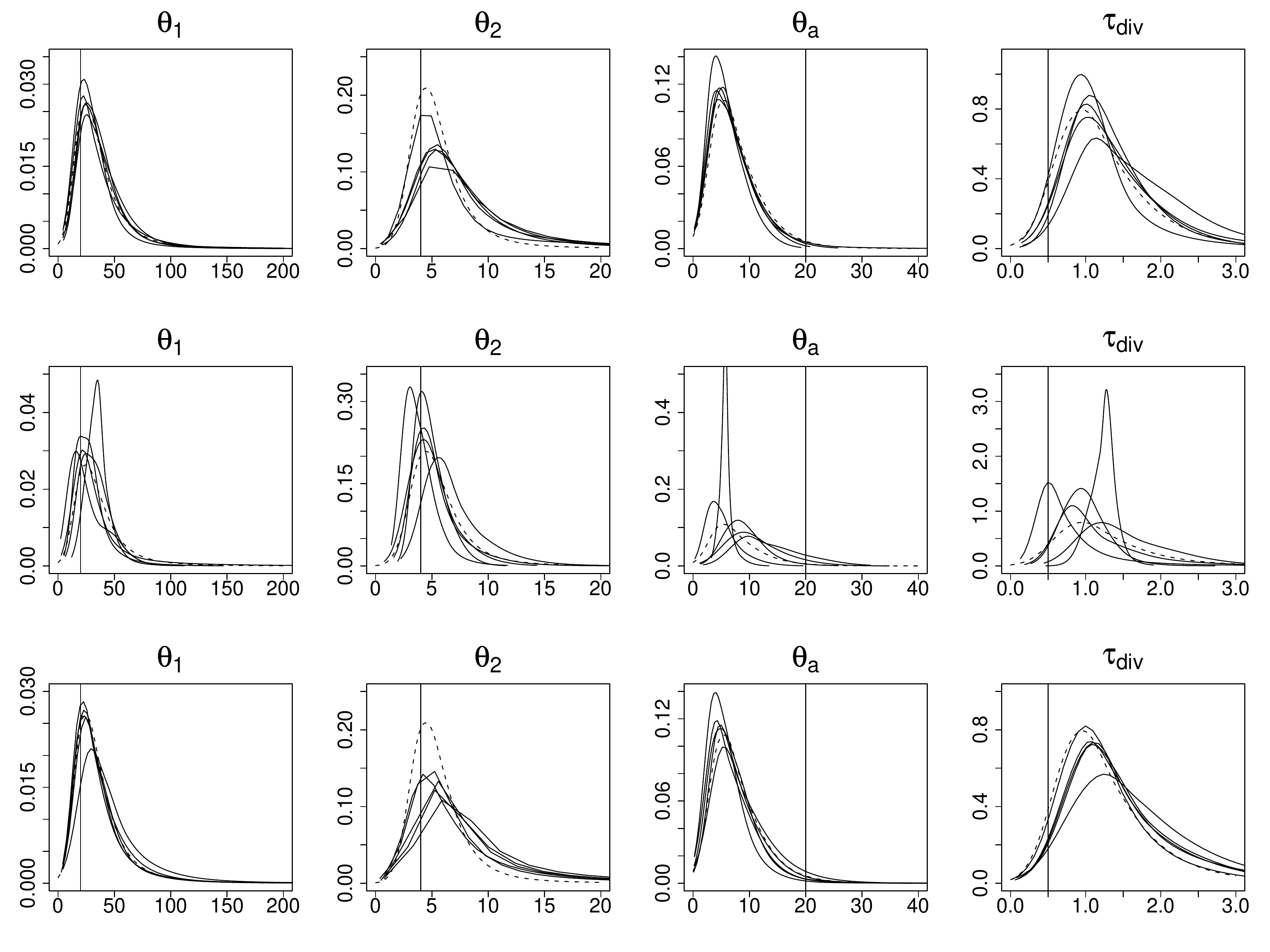}}
\caption{Variability of the different alternative schemes evaluated though five replicates
of the density estimates of the posterior distributions of four identifiable
parameters of the population genetics model. First line: population Monte Carlo
version; second line: tempered Markov chain Monte Carlo version; third line: Beaumont
et al.'s (2002) version of the approximate Bayesian computation algorithm, against the 
reference posterior, obtained by $5\times10^5$ simulations in Beaumont et al.'s (2002) version
(dotted line). The vertical lines identify the true values of the parameters.
}\label{fig:popgenvar}
\end{figure}

In order to operate a fair comparison between the three methods, the computing times
are kept identical. Thus, reference posterior distributions have been constructed with 
Beaumont et al.'s (2002) approach, 
using $N=3\times 10^4$ simulated datasets and choosing $\epsilon$ as the $0\cdot01$ quantile of the distances. 
For Bortot et al.'s (2007) approach, the tuning parameter driving the acceptance threshold $\epsilon$ is
drawn from an exponential prior with parameter equal to the mean distance of the ten closest datasets (among $5000$
generated in a pilot simulation), followed by $23,000$ regular iterations with a thinning factor of $23$.
Each Markov move combined two simultaneous updates : an independent draw of $\epsilon$ from its exponential prior and a
random walk based on a log-normal deviate with standard deviation equal to $0\cdot3$ of one of the other parameters chosen uniformly 
at random, the latter involving the
computation of a Hastings term. Lastly, in the population Monte Carlo version, $\epsilon_1$ is based on
the preliminary simulation as the $0\cdot1$ quantile and four iterations are performed with 
$\epsilon_2=0\cdot75\epsilon_1$, $\epsilon_3=0\cdot9\epsilon_2$ 
and $\epsilon_4=0\cdot9\epsilon_3$, and with truncated normals in the random walk. 
Each iteration is based on samples of size $10^3$. All three versions then require an average $4\cdot5$ minutes.
Figure \ref{fig:popgenvar} summarises the output of the comparison between the three methods 
over five independent runs for the same simulated dataset and 
it shows that, at least in this example, the posterior evaluations based on population Monte Carlo are 
more stable than Bortot et al.'s (2007) algorithm, while comparable with Beaumont et al.'s (2002).

\section{Conclusion}
While Sisson et al.'s (2007) algorithm induces biased weights, with a visible impact on the quality of
the approximation, we have shown that the same Markov transition kernels and thus the same 
computing power can be used to produce an unbiased scheme. 

The population Monte Carlo scheme is based on an importance argument that does not require a backward kernel 
as in \eqref{eq:sissonPRC22}. We have thus established that the adaptive scheme of 
\cite{cappe:douc:guillin:marin:robert:2007} is also appropriate in this setting, towards a better fit
of the proposal kernel $K_t$ to the target $\pi\{ \theta\mid \varrho(x,y)<\epsilon\}$. From a practical point
of view, the number of iterations $T$ can be controlled via the modifications 
in the parameters of $K_t$, a stopping rule being that the iterations should stop when those parameters 
have settled, while the more fundamental issue of selecting a sequence of $\epsilon_t$'s towards a 
proper approximation of the true posterior can rely on the stabilisation of the estimators of some
quantities of interest associated with this posterior.

\section*{Acknowledgements}
This research is partly supported by the Agence Nationale de la Recherche through the Misgepop project.
Both last authors are affiliated with CREST, Paris.
Parts of this paper were written by the last author in the Isaac Newton Institute,
Cambridge, whose peaceful and stimulating environment was deeply appreciated. 
Helpful comments from the editorial board of {\em Biometrika} and from O.~Capp\'e are 
gratefully acknowledged.

\bibliography{biblio}
\bibliographystyle{biometrika}
\end{document}